%% file: bare_jrnl_new_sample4.tex
\documentclass[lettersize,journal]{IEEEtran}
\usepackage{amsmath,amsfonts}
\usepackage{amssymb}
\usepackage[ruled,vlined]{algorithm2e}
\usepackage{array}
\usepackage[caption=false,font=normalsize,labelfont=sf,textfont=sf]{subfig}
\usepackage{textcomp}
\usepackage{stfloats}
\usepackage{url}
\usepackage{verbatim}
\usepackage{graphicx}
\usepackage{cite}
\usepackage{cleveref}
\include{Wgroup-Teaching-Preamble}

\usepackage{acronym}
\acrodef{co}[CO]{combinatorial optimization}
\acrodef{nco}[NCO]{neural combinatorial optimization}
\acrodef{rp}[RP]{routing problem}
\acrodef{rl}[RL]{reinforcement learning}
\acrodef{tsp}[TSP]{traveling salesman problem}
\acrodef{tsptw}[TSPTW]{traveling salesman problem with time window}
\acrodef{vrp}[VRP]{vehicle routing problem}
\acrodef{cvrp}[CVRP]{capacitated vehicle routing problem}
\acrodef{vrptw}[VRPTW]{vehicle routing problem with time window}
\acrodef{gnn}[GNN]{graph neural network}
\acrodef{gam}[GAM]{graph attention model}
\acrodef{gcn}[GCN]{graph convolutional network}
\acrodef{egam}[EGAM]{extended graph attention model}
\acrodef{dl}[DL]{deep learning}
\acrodef{nn}[NN]{neural network}
\acrodef{drl}[DRL]{deep reinforcement learning}
\acrodef{sota}[SOTA]{state-of-the-art}
\acrodef{or}[OR]{operations research}
\acrodef{ai4rp}[AI4RP]{Artificial Intelligence for Routing Problem}
\acrodef{pn}[PN]{Pointer Network}
\acrodef{s2v}[S2V]{Structure2Vec}
\acrodef{cnn}[CNN]{convolutional neural network}
\acrodef{mha}[MHA]{multi-head attention}
\acrodef{ml}[ML]{machine learning}
\acrodef{sl}[SL]{supervised learning}
\acrodef{ff}[FF]{feed-forward}
\acrodef{ffn}[FFN]{feed-forward network}
\acrodef{rnn}[RNN]{recurrent neural network}
\acrodef{nlp}[NLP]{natural language processing}
\acrodef{egat}[EGAT]{edge-featured GAT}
\acrodef{pip}[PIP]{Proactive Infeasibility Prevention}
\acrodef{tspdl}[TSPDL]{traveling salesman problem with draft limit}
\acrodef{pctsp}[PCTSP]{prize collecting traveling salesman problem}
\acrodef{llm}[LLM]{large language model}
\acrodef{vit}[ViT]{vision transformer}

\acrodef{mrta}[MRTA]{multi-robot task allocation}
\acrodef{mvrp}[MVRP]{multi-vehicle routing problem}
\acrodef{cbba}[CBBA]{Consensus-Based Bundle Algorithm}
\acrodef{acbba}[ACBBA]{Asynchronous \ac{cbba}}
\acrodef{pi}[PI]{performance impact}
\acrodef{ipi}[IPI]{inclusion performance impact}
\acrodef{rpi}[RPI]{removal performance impact}
\acrodef{pomo}[POMO]{Policy Optimization with Multiple Optima}
\acrodef{am}[AM]{Attention Model}
\acrodef{marl}[MARL]{multi-agent reinforcement learning}
\acrodef{ccn}[CCN]{covariant compositional network}
\acrodef{mdgam}[MDGAM]{multi-decoder graph attention model}
\acrodef{grmapg}[GRMAPG]{group relative multi-agent policy gradient}
\acrodef{mdp}[MDP]{Markov Decision Process}
\acrodef{decpomdp}[Dec-POMDP]{Decentralized Partially Observable Markov Decision Process}
\acrodef{ppo}[PPO]{Proximal Policy Optimization}
\acrodef{mappo}[MAPPO]{multi-agent proximal policy optimization}
\acrodef{maa2c}[MAA2C]{multi-agent advantage actor-critic}
\acrodef{ctde}[CTDE]{centralized training and decentralized execution}
\acrodef{gat}[GAT]{graph attention network}

\usepackage{amsthm}
\theoremstyle{plain}

\theoremstyle{definition}

\theoremstyle{remark}
\newtheorem*{remark}{Remark}

\usepackage{multirow}
\usepackage{booktabs} 

\begin{document}

\title{MDGAM-Based Cooperative Task Scheduling for Communication-Constrained Distributed Multi-Agent Systems}

\author{Licheng Wang, Mingtao Huang, Yuan Shen
\thanks{The authors are with the Department of Electronic Engineering, and Beijing National Research Center for Information Science and Technology, Tsinghua
University, Beijing 100084, China.}
\thanks{Arxiv preprint.}}

\markboth{Journal of \LaTeX\ Class Files,~Vol.~14, No.~8, August~2021}%
{Shell \MakeLowercase{\textit{et al.}}: A Sample Article Using IEEEtran.cls for IEEE Journals}


\maketitle

\begin{abstract}

Cooperative task scheduling in communication-constrained distributed multi-agent systems is challenging because each agent must make decisions from partial and dynamic observations while satisfying complex practical constraints. Existing heuristics rely on handcrafted bidding rules and repeated consensus, whereas many learning-based methods assume global observations and lack explicit communication-based coordination. To address these limitations, this paper proposes a neural scheduling framework for distributed \ac{mrta}, consisting of a \ac{mdgam} policy model and a critic-free \ac{grmapg} training algorithm. \Ac{mdgam} uses an extended graph attention mechanism to jointly update node and edge features, and employs multiple decoders to generate task-selection decisions and communication messages. \Ac{grmapg} constructs group-relative advantages from equivalent task-planning instances to replace the critic network used in conventional \ac{marl} algorithms, thereby reducing training difficulty and improving convergence performance. Experiments under different problem scales and communication ranges show that the proposed method improves task-completion performance over existing heuristic and learning-based methods, while ablation, complexity, and generalization tests further validate the proposed innovations.

\end{abstract}

\def\abstractname{Note to Practitioners}
\begin{abstract}
This work is motivated by cooperative task scheduling in distributed multi-agent applications such as intelligent transportation, Internet of Things, and emergency response, where robots, vehicles, or UAVs must execute tasks in a distributed manner under limited communication. In these scenarios, heterogeneous agents and complex task constraints require efficient inter-agent coordination to improve execution efficiency. Unlike existing learning-based methods that mainly focus on individual route construction or rely on global observations, the proposed framework explicitly considers both task planning and communication-based coordination among agents. This design enables agents to improve local planning while sharing and utilizing useful coordination information. Experimental results show that the proposed method achieves higher task-completion performance than existing distributed algorithms, with additional advantages in runtime and communication cost compared with classical auction-based algorithms. The framework is suitable for a broad range of distributed task-planning applications and can be flexibly transferred to related scenarios with different practical constraints and settings.
\end{abstract}
\def\abstractname{Abstract}


\acresetall

\section{Introduction}
\IEEEPARstart{D}{istributed} multi-agent systems are increasingly deployed in intelligent transportation, Internet of Things, and emergency-response scenarios, where robots, vehicles, or UAVs must cooperate to serve spatially distributed tasks under limited sensing, computation, and communication resources \cite{LiuLiWu:J20,ShoHakWas:J23,SuLin:J15,BaiFieKro:J22}. \Ac{mrta} addresses the assignment and scheduling of tasks for a robot team through centralized optimization or distributed coordination \cite{KhaHusElm:J15,ZhaZhaLam:J23}. Given heterogeneous robots and spatially distributed tasks, its objective is to determine task-execution sequences that optimize a global performance metric, such as task completion, collected reward, or travel cost \cite{CheSun:J12,GenCheMgu:J21,CheAloBai:J21}.


\begin{figure}[!t]
  \centering
  \includegraphics[width=\columnwidth]{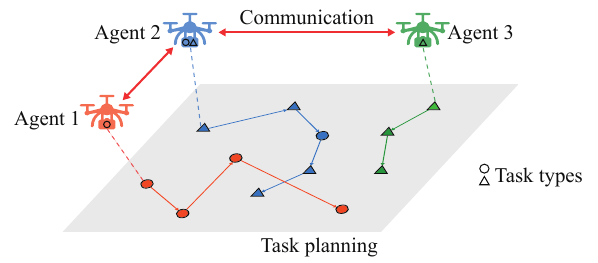}
  \caption{Illustration of multi-agent cooperative task planning. Two task types are shown as an example, each of which must be completed by an agent with the corresponding capability. Agents within communication range can exchange information to improve task-completion efficiency.}
  \label{fig:problem scenario}
\end{figure}

From an optimization perspective, \ac{mrta} is a combinatorial optimization problem closely related to variants of \ac{mvrp}, and is generally NP-hard \cite{GerMat:J04}. As the numbers of robots and tasks increase, the associated search space grows rapidly, making exact optimization difficult for large-scale instances \cite{LiuWeiWan:J21}. In practical scenarios, constraints such as time windows make task feasibility highly dependent on each task's position in the planned sequence, thereby requiring more carefully coordinated task-planning strategies \cite{JiaYuAi:J18}. The difficulty becomes more pronounced in distributed settings, where each robot only has access to partial and time-varying information and must coordinate with nearby peers through limited communication links \cite{ChaHonWan:J14,YanShe:J24}. In this paper, we consider a communication-constrained \ac{mrta} problem with heterogeneous robot attributes, various task constraints, and limited communication ranges. Under this setting, the inter-agent communication topology is time-varying and may be divided into several disconnected components. We aim to develop a policy that jointly accounts for communication coordination and task planning under such conditions, so as to maximize the total number of completed tasks. Fig.~\ref{fig:problem scenario} illustrates a cooperative task-planning scenario with task capabilities and range-limited communication.

Existing approaches to \ac{mrta} can be broadly categorized into handcrafted heuristics and learning-based methods \cite{ChoBruHow:J09,XuFanChe:J21}. Among heuristic methods, \ac{cbba} \cite{BruChoHow:C08} and \ac{pi}-based algorithms \cite{ZhaMenChu:J15} are two representative families. \Ac{cbba} constructs task bundles according to bidding functions and resolves conflicting assignments through decentralized consensus \cite{CheLiChe:J25,BaiYanGe:J21}. \Ac{pi}-based methods use performance-impact metrics to evaluate task insertion and removal effects. PI-maxAss \cite{TurMenSch:J17} improves task completion through distributed rescheduling, EEPI \cite{WanLiuQiu:J23} reduces the required iterations through improved metric design, and GPI \cite{WanLiHua:J25} adopts leader-follower grouping for communication. However, these heuristic methods rely on handcrafted bidding or performance-impact functions and require repeated replanning under communication constraints and time-varying local observations, resulting in considerable communication and computational overhead.

Learning-based methods have become increasingly effective for solving \ac{co} problems, especially \ac{vrp}-related variants. \Ac{rl}-based \ac{nco} methods learn solution-construction policies directly from optimization objectives, reducing the reliance on handcrafted heuristics or labeled datasets. Representative studies have developed neural routing policies based on Pointer Networks \cite{VinForJai:C15,BelPhaLeNorBen:A16}, \ac{gcn} \cite{KhaDaiZhaDilSon:C17,JosLauBre:A19}, and \ac{am} \cite{KoovanWel:C19, JinDingPan:A23}, while later improvements further enhance training, inference, and constraint handling for single-vehicle routing problems \cite{KwoChoKimYooGwoMin:C20,BiMaZho:C24}.

Several recent studies have extended learning-based approaches to multi-vehicle or multi-agent settings. MAPDP \cite{ZonZheLi:C22} introduced a centralized \ac{marl} framework for pickup-and-delivery planning, where pickup-delivery dependencies and inter-vehicle cooperation are captured via paired embeddings and cooperative decoders. DL-DRL \cite{MaoWuFan:J24} employed a hierarchical two-level architecture for multi-UAV scheduling, in which the upper level performs task allocation and the lower level determines the execution sequence for each UAV. The hierarchical DQN-based method in \cite{WanLiShe:J25} decomposes the problem into target assignment and resource allocation, focusing on efficient matching between tasks and resources, but does not optimize the task execution sequences.
CAM \cite{PauCho:J24} learns distributed \ac{mrta} policies over task graphs using a \ac{ccn}-based encoder and an attention-based decoder.
However, CAM assumes full observability and does not account for time-varying communication topology. 
Therefore, the applicability of these learning-based multi-agent methods to communication-constrained distributed task scheduling remains limited.

In this work, we propose a neural scheduling framework for communication-constrained distributed \ac{mrta}. The problem is formulated with task temporal constraints, heterogeneous agent attributes, and limited communication ranges, and the distributed decision process is described under the \ac{decpomdp} framework. We design \ac{mdgam} as a multi-output policy module with a task decoder and a communication decoder, supported by an extended graph attention encoder for graph-structured input representation. For policy optimization, we develop \ac{grmapg}, a critic-free training algorithm that constructs group-relative advantages from equivalent task-planning instances, reducing trainable parameters and improving convergence for the terminal task-completion objective.

Experiments show that the proposed method achieves higher task-completion performance than existing heuristic and learning-based algorithms. Under the medium-scale setting with an intermediate communication range, it improves the number of completed tasks over PI-maxAss and CAM by $4.13\%$ and $3.74\%$, respectively. Additional ablation, complexity, and generalization tests further evaluate the proposed method in terms of component effectiveness, runtime and communication cost, and robustness.

The main contributions of this work are summarized as follows:
\begin{itemize}
    \item We formulate a communication-constrained distributed task scheduling problem with various task and robot constraints. The decentralized decision process is modeled as a \ac{decpomdp}, where each robot plans from partial observations and exchanged messages to maximize the shared objective of total task completion.
    \item We propose \ac{mdgam}, an edge-aware multi-decoder graph attention architecture for distributed task planning. \Ac{mdgam} uses an extended graph attention encoder to represent local observation graphs with both node and edge features, and employs a task decoder and a communication decoder to jointly generate task decisions and inter-agent messages.
    \item We develop \ac{grmapg}, a critic-free training algorithm that constructs advantages from equivalent task-planning instances. Compared with conventional \ac{marl} algorithms, \ac{grmapg} reduces the number of trainable parameters and improves training convergence while enabling end-to-end policy optimization without ground truth.
    \item We conduct comprehensive experiments covering task-completion performance, complexity, ablation analysis, and generalization capability. The proposed method outperforms representative heuristic and learning-based baselines, and additional ablation studies verify the individual contributions of the proposed model and training algorithm.
\end{itemize}

The remainder of this paper is organized as follows. Section II formulates the communication-constrained task scheduling problem and describes its \ac{decpomdp} representation. Section III presents the proposed \ac{mdgam} architecture, from the extended graph attention mechanism to the graph-embedding encoder and the multiple decoders. Section IV presents actor-critic training with an attention-based critic and then introduces the proposed \ac{grmapg} training algorithm. Section V evaluates the proposed method through performance comparisons, ablation studies, complexity analysis, and generalization tests. Finally, Section VI concludes the paper.

\section{Problem Formulation}
\subsection{Problem settings}

In this paper, we investigate a task planning problem in communication-constrained distributed multi-agent systems. Consider a scenario in which multiple locations within a given area require autonomous task execution. The tasks belong to different types, and each task must be completed by an agent equipped with the corresponding capability. In addition, task execution is subject to both time-window constraints and service-duration requirements. Due to communication limitations, only agents within communication range can exchange information. Under this setting, our goal is to distributively determine the task sequence of each agent so as to maximize the total number of completed tasks. The problem is formulated as follows.

The task set is denoted by ${\Set T} = \{1,2,\dots,N\}$, and each task $n\in{\Set T}$ is characterized by its location $\V{p}_{n}\in{\mathbb R}^2$, time window $[\underline{t}_n, \overline{t}_n]$, service time $\tau_n$, and task type $\kappa_n\in\{1,2,\dots,C\}$, where $C$ denotes the number of task types. The time-window constraint specifies that an agent may initiate a task only if it arrives at the task location within the prescribed time window, after which the agent spends the required service time to complete the task. If the agent arrives before the opening of the time window, it must wait until service can begin. The task-type constraint indicates that a task can be completed only by an agent possessing the required capability.

The agent set is denoted by ${\Set R} = \{1,2,\dots,M\}$, where an agent can represent a robot, a UAV, or another autonomous platform. Each agent is associated with the following attributes: depot location $\V{p}_{m}^{0}\in{\mathbb R}^2$, task capability ${\Set K}_m\subseteq\{1,2,\dots,C\}$ with ${\Set K}_m\neq \emptyset$, travel speed $v_m$, and latest return time $T_{\rm r}$. Each agent initially departs from its depot, visits the assigned task locations, and finally returns to the depot no later than its latest return time. An agent may possess the capabilities required for one or multiple task types. In addition, two agents can directly communicate only when their distance is within a communication radius $r$.

Consistent with common distributed \ac{mrta} formulations \cite{TurMenSch:J17,WanLiuQiu:J23,BaiYanGe:J21}, communication limitations are modeled at the topology level. The communication radius defines a time-varying communication graph, and information exchange is allowed only within each communication-connected component; low-level network effects such as bandwidth, delay, and packet loss are not explicitly modeled.

The decision variable is the task sequence of each agent, denoted by $\V{\sigma}_m=(\sigma_{m,1}, \sigma_{m,2}, \dots, \sigma_{m,|\V{\sigma}_m|})$, where $\sigma_{m,i}\in{\Set T}$. A task included in a sequence is considered feasible only if both the time-window constraint and the capability constraint are satisfied. Let $t_{{m,i}}^{\rm arr}$ and $t_{{m,i}}^{\rm dep}$ denote the arrival time and departure time of agent $m$ at task $\sigma_{m,i}$, respectively. Then, the temporal relationships associated with agent $m$ executing the task sequence $\boldsymbol{\sigma}_m$ are given by
{
\begin{align}
&t_{{m,i}}^{\rm arr}=\begin{cases}d(\sigma_{m,i},\V{p}^0_m)/v_m& i=1\\
d(\sigma_{m,i},\sigma_{m,i-1})/v_m+t_{{m,i-1}}^{\rm dep}&\text{otherwise}\end{cases},\\
&t_{{m,i}}^{\rm dep}=\begin{cases}\tau_{\sigma_{m,i}}+\max(t_{{m,i}}^{\rm arr},\underline{t}_{\sigma_{m,i}})&\sigma_{m,i}\text{ is feasible}\\
t_{{m,i}}^{\rm arr}&\text{otherwise}\end{cases}
.
\end{align}
}%
In addition, if a task conflict occurs, that is, if more than one agent selects the same task, the task is regarded as invalid for the agent that reaches it later.

From a centralized perspective, the global optimization problem can be formulated as
\begin{subequations}
\begin{align}
\mathcal{P}: \quad\max_{\V{\sigma}_m, m\in{\Set R}} \quad & \sum_{m=1}^{M}|\V{\sigma}_m|& \\
\mbox{s.t.}\quad
&t_{{m,i}}^{\rm arr}\leq\overline{t}_{\sigma_{m,i}}, & \label{constraint_ddl}\\
&\kappa_{\sigma_{m,i}}\in{\Set K}_m, & \label{constraint_cap}\\
&\V{\sigma}_{m}\cap\V{\sigma}_{m'}=\emptyset, & \label{constraint_conflict}\\
&t_{{m,|\V{\sigma}_m|}}^{\rm dep}+\frac{d(\sigma_{m,|\V{\sigma}_m|},\V{p}^0_m)}{v_m}\leq T_{\rm r},&  \label{constraint_rett}\\
& \forall m,m'\in{\Set R}, m\neq m',  i\in \{1,2,\ldots,|\V{\sigma}_m|\}.& \nonumber
\end{align}
\end{subequations}
where \eqref{constraint_ddl}--\eqref{constraint_rett} represent the constraints on the time window, agent capability, conflict avoidance, and return time, respectively. The problem $\mathcal{P}$ describes the full-observation centralized objective and serves only as a reference, since individual agents cannot directly solve it in the distributed setting. Moreover, $\mathcal{P}$ is NP-hard and involves a rapidly growing search space of coupled task sequences. Therefore, each agent must plan through decentralized sequential decision-making based on local observations and information exchanged within its communication-connected neighborhood. The next subsection models this process as a \ac{decpomdp} \cite{BerGivImm:J02}.

\subsection{Distributed Decision-Process Modeling}
\label{subsec:decpomdp}

We describe the distributed task planning process using a \ac{decpomdp} \cite{BerGivImm:J02}, defined by $\left({\Set S}, \{{\Set A}_m\}, P, R, \{{\Set O}_m\}, O, \gamma\right)$. Here, ${\Set S}$ is the global state space, ${\Set A}_m$ and ${\Set O}_m$ are the local action and observation spaces of agent $m$, $P$ and $O$ are the state-transition and observation functions, $R$ is the shared team reward, and $\gamma$ is the discount factor. Each agent acts on locally available information under communication constraints, while all agents share the objective of maximizing the total number of completed tasks over the episode.

The task set and static task attributes are available to all agents before execution as mission-level information. The decision process remains partially observable because each agent still plans online based on its own evolving state and the dynamic information received through the current communication topology, while the execution states of disconnected agents and task-conflict information may be unavailable.

The decision step $j$ in this subsection denotes an event-triggered global decision index rather than a synchronized global round, while $i$ is used later to index the local task-selection sequence of an individual agent. At the initial time, each agent makes its first task-selection decision. Afterwards, an agent triggers a new decision event only after it has traveled to the selected task, waited if necessary, and completed the required service. Agents that are traveling or serving a task do not generate new task-selection actions, while other agents may continue to trigger their own decision events according to their local completion times. Therefore, the distributed decision process evolves asynchronously, and each agent maintains its own local time, current position, planned task sequence, and latest generated communication message.

At each decision event, the communication topology is determined by the currently available agent positions and the communication range. The active agent collects the states and stored messages of agents that are reachable through the current communication-connected component, and then constructs its local observation. The policy outputs both a task-selection action and a new communication message. The selected task is checked against capability, temporal and conflict constraints before being added to the agent's planned sequence, while the newly generated message is stored by the active agent and can be transmitted to other agents at later decision events when they are communication-reachable. Thus, communication is restricted by the time-varying topology, although multi-hop information propagation within a connected component is allowed before each decision event.

Let $\V{s}^j$ denote the global state at decision event $j$, and let $m_j$ denote the active agent at this event. The global state evolves according to
\begin{equation}
\V{s}^{j+1}\sim P(\V{s}^{j+1}\mid \V{s}^j,a_{m_j}^j, \RV{c}_{m_j}^j),
\end{equation}
where $a_{m_j}^j$ is the task-selection action generated by the active agent. At each decision event, the active agent selects its next task based on its local observation using a policy parameterized by $\V{\theta}$:
\begin{equation}
a_{m_j}^j \sim P_{\V{\theta}}(a_{m_j}^j\mid \V{o}_{m_j}^j).
\label{task_selection_policy}
\end{equation}
Here, the policy outputs a selection-probability distribution over candidate tasks and the depot. Infeasible choices are masked out according to the capability, time-window, conflict, and return-time constraints.
To support inter-agent coordination, the policy additionally outputs a communication message generated from the local observation as
\begin{equation}
\RV{c}_{m_j}^j=f_{\V{\theta}}^{\rm c}(\V{o}_{m_j}^j),
\label{communication_message_policy}
\end{equation}
where $f_{\V{\theta}}^{\rm c}$ denotes the communication-message mapping. The communication message in \eqref{communication_message_policy} is stored by the active agent and can be incorporated into subsequent local observations of communication-reachable agents.

At each decision event $j$, the local observation $\V{o}_{m_j}^j$ of the active agent consists of the following components:
\begin{itemize}
  \item The agent's own information, including its current position $\V{p}_{m_j}^{\rm cur}$, current local time $t$, completed task list $\xi_{m_j}\subseteq{\Set T}$, and intrinsic attributes $\V{p}^0_{m_j}$, ${\Set K}_{m_j}$, $v_{m_j}$, and $T_{\rm r}$;
  \item The static information of all tasks $n\in{\Set T}$, including their attributes $\V{p}_n$, $[\underline{t}_n, \overline{t}_n]$, $\tau_n$, and $\kappa_n$;
  \item The dynamic information of other agents in the current communication-connected component. Let ${\Set N}_{m_j}^j$ denote the set of agents that can exchange information with agent $m_j$ through local message propagation at decision event $j$. For each $m'\in{\Set N}_{m_j}^j$, agent $m_j$ can access its available current state, including $\V{p}_{m'}^{\rm cur}$ and $\xi_{m'}$, as well as its intrinsic attributes $\V{p}^0_{m'}$, ${\Set K}_{m'}$, and $v_{m'}$, and can further receive its latest stored message $\RV{c}_{m'}^{\rm last}$. Agents outside this component are unobservable to agent $m_j$ at this event.
\end{itemize} 
Before each decision event, agents within the same communication-connected component are assumed to complete a local information-exchange phase. This abstraction corresponds to multi-hop propagation of available agent states and learned messages within the component. Agents belonging to different disconnected components cannot exchange information at that event. Therefore, each agent plans from a partial and time-varying local view rather than from the full system state. An episode terminates when no agent can execute any further feasible task and all agents have either returned to their depots or no further feasible decisions remain.

Given an instance $\V{g}$ and the joint task-selection trajectory $\V{\sigma}=(\V{\sigma}_1,\V{\sigma}_2,\ldots,\V{\sigma}_M)$, the terminal reward $R(\V{\sigma},\V{g})$ is defined as the total number of tasks completed over the episode. The policy optimization objective is therefore given by
\begin{equation}
\max_{\V{\theta}}\mathbb{E}_{\V{g}\sim P(\V{g}),\V{\sigma}\sim P_{\V{\theta}}(\V{\sigma}|\V{g})}\left[R(\V{\sigma},\V{g})\right].
\label{policy_objective}
\end{equation}
Here $P_{\V{\theta}}(\V{\sigma}|\V{g})$ denotes the probability that the policy generates the joint trajectory $\V{\sigma}$. Under decentralized execution, this probability factorizes over agents and their local decision indices conditioned on their local observations:
\begin{equation}
P_{\V{\theta}}(\V{\sigma}|\V{g})=\prod_{m=1}^M\prod_{i=1}^{|\V{\sigma}_m|}P_{\V{\theta}}(a_m^i=\sigma_{m,i}|\V{o}_m^i)
.
\label{trajectory_probability}
\end{equation}
To parameterize this decentralized policy, the next section introduces a graph-based neural architecture that encodes each agent's local observation and generates both task-selection decisions and communication messages.

\section{Proposed Network Architecture}
\begin{figure*}[ht]
\begin{minipage}[c]{1.0\linewidth}
  \centering
  \centerline{\includegraphics[width=0.98\textwidth]{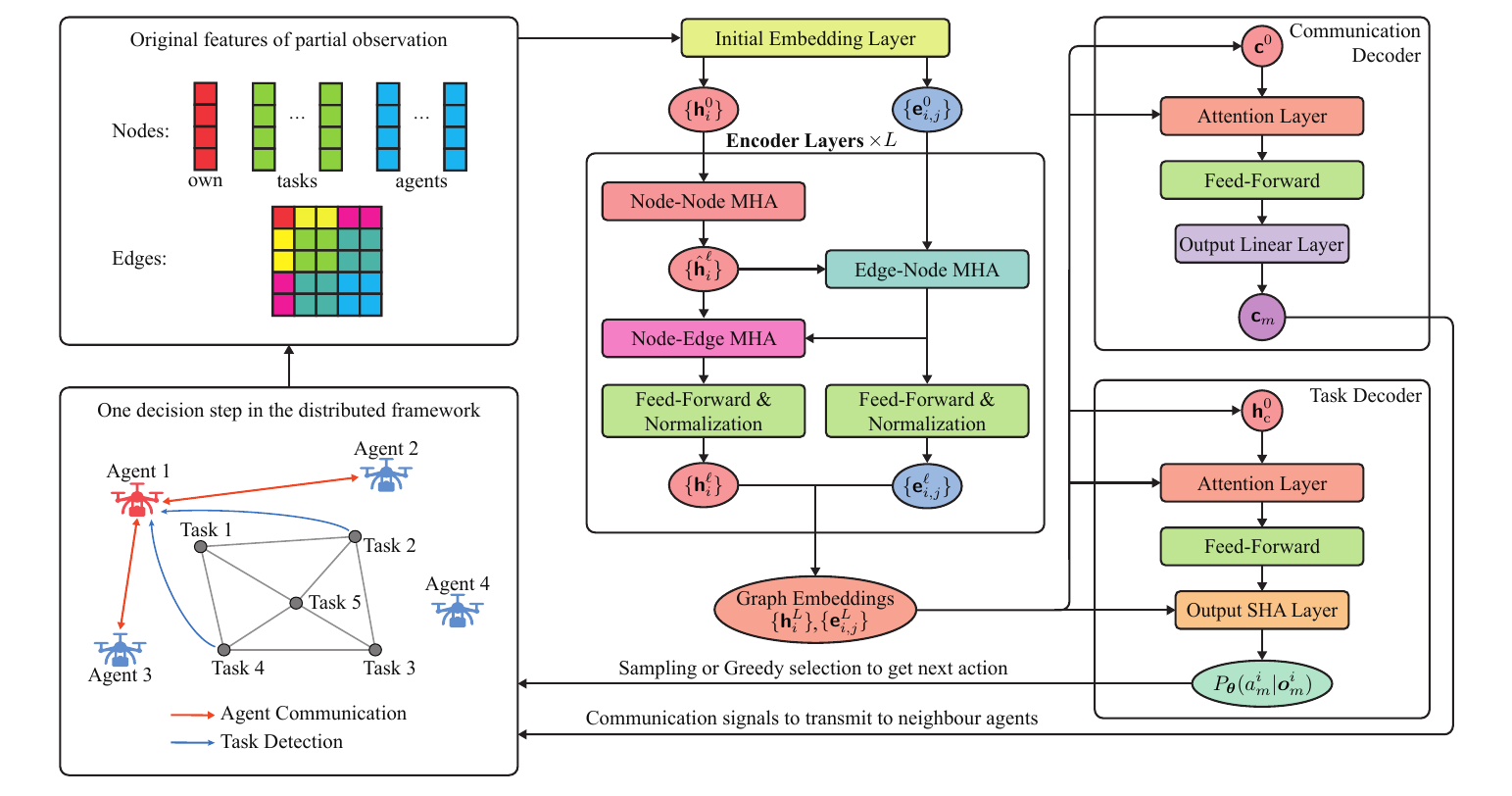}}
\end{minipage}
\caption{Architecture of the proposed \ac{mdgam} for one decentralized decision step. The rounded rectangles represent trainable network layers, while the ellipses denote intermediate and output variables. At each step, the current agent, highlighted in red, constructs a graph-structured local observation from candidate tasks and neighboring agents, which is then provided to \ac{mdgam} in the form of node and edge features. The arrows indicate the data flow from the step input to task selection and message generation. After one step is completed, the environment state is updated, and the next decision step repeats this process until the episode terminates.}
\label{fig:main model}
\end{figure*}

This section presents the proposed \ac{mdgam} for decentralized policy generation. As illustrated in Fig.~\ref{fig:main model}, \ac{mdgam} takes the graph-structured local observation of each agent as input and produces two outputs: a task-selection distribution and a communication message for inter-agent coordination. The section first introduces the extended graph attention mechanism as the basic representation-learning operation. It then describes the encoder that embeds the local observation graph. Finally, the task decoder and the communication decoder are presented to generate task decisions and communication messages, respectively.
  
\subsection{Extended graph attention mechanism}
The local observation available to each agent has a natural graph structure: nodes represent the agent itself, its depot, candidate tasks, and neighboring agents, while edges encode pairwise relations such as travel distance and communication connectivity. For communication-constrained task scheduling, node-only representations are insufficient because feasibility and coordination decisions are strongly affected by edge-dependent factors, including travel time, time-window compatibility, and inter-agent communication links. Therefore, \ac{mdgam} employs an extended graph attention mechanism that jointly updates node and edge representations. This mechanism follows the general idea of edge-aware attention in \cite{WanYanHua:A26}, but is adapted here to represent the local observation graphs used for decentralized task planning.

The proposed block is built upon the multi-head scaled dot-product attention operation \cite{VasShaPar:C17}. Let the input query, key, and value features be denoted by $\RM{X}=[\RV{x}_1,\RV{x}_2,\cdots,\RV{x}_M]$, $\RM{Y}=[\RV{y}_1,\RV{y}_2,\cdots,\RV{y}_N]$, and $\RM{Z}=[\RV{z}_1,\RV{z}_2,\cdots,\RV{z}_N]$, respectively. The MHA operation is defined as
{
\begin{align}
\operatorname{MHA}(\RM{X}, \RM{Y},\RM{Z})=\M{W}^{\rm O}\operatorname{Concat}(\RM{H}_1,\RM{H}_2,\ldots,\RM{H}_h),\\
\RM{H}_i = \operatorname{Attention}(\RM{X}^{\rm T}\M{W}^{\rm Q}_i,\RM{Y}^{\rm T}\M{W}^{\rm K}_i,\RM{Z}^{\rm T}\M{W}^{\rm V}_i)^{\rm T}.
\end{align}}%
where the attention operation is given by
\begin{equation}\label{attention}
\operatorname{Attention}(\RM{Q},\RM{K},\RM{V})=\operatorname{softmax}\left(\frac{\RM{Q} \RM{K}^{\rm T}}{\sqrt{d_{\rm k}}}\right) \RM{V}.
\end{equation}
In \eqref{attention}, $d_{\rm k}$ denotes the last dimension of $\RM{K}$; when $\RM{Y}=\RM{Z}$, we write $\operatorname{MHA}(\RM{X}, \RM{Y})$ for brevity. 

Given node representations $\{\RV{h}_i^{\ell-1}\}$ and edge representations $\{\RV{e}_{ij}^{\ell-1}\}$ at layer $\ell-1$, the extended graph attention mechanism updates them through three attention operations. First, Node-Node Attention aggregates information from all observable nodes and produces an intermediate node representation:
\begin{equation}
\hat{\RV{h}}_i^{\ell}=\operatorname{MHA}(\RV{h}_i^{\ell-1}, [\RV{h}_1^{\ell-1},\RV{h}_2^{\ell-1},\cdots,\RV{h}_N^{\ell-1}]).
\label{nna}
\end{equation}
Node-Node Attention is effective for extracting node-level contextual features \cite{KoovanWel:C19,JinDingPan:A23}, but it does not explicitly update pairwise relations. This motivates the following two edge-aware attention operations.

Second, Edge-Node Attention updates each edge representation using the representations of its incident nodes. Since the edge from node $i$ to node $j$ may differ from the edge from node $j$ to node $i$, we treat edges as directed and introduce a positional encoding operator $\operatorname{PE}(\cdot,\cdot)$ to distinguish the start and end nodes:
\begin{equation}
\RV{e}_{i,j}^{\ell}=\operatorname{MHA}(\RV{e}_{ij}^{\ell-1}, \operatorname{PE}(\hat{\RV{h}}_i^{\ell},\hat{\RV{h}}_j^{\ell})).
\label{ena2}
\end{equation}

\begin{remark}
The positional encoding operator $\operatorname{PE}(\cdot,\cdot)$ can be implemented in different forms, such as fixed sinusoidal encodings or learnable embedding layers \cite{VasShaPar:C17,DevChaLee:C19}. In this paper, we adopt a lightweight direction-aware encoding. Specifically, the representation of the start node is concatenated with a scalar tag $-1$, while the representation of the end node is concatenated with a scalar tag $1$, i.e.,
\begin{equation}
\operatorname{PE}(\hat{\RV{h}}_i^{\ell},\hat{\RV{h}}_j^{\ell})
=
[\operatorname{Concat}(\hat{\RV{h}}_i^{\ell},-1),\operatorname{Concat}(\hat{\RV{h}}_j^{\ell},1)].
\end{equation}
This simple design enables the attention module to distinguish the start and end nodes of a directed edge without introducing additional trainable parameters.
\end{remark}

Third, Node-Edge Attention propagates edge information back to the node representations:
\begin{equation}
\RV{h}_{i}^{\ell}=\operatorname{MHA}(\hat{\RV{h}}_{i}^{\ell}, [\RV{e}_{i,1}^{\ell},\RV{e}_{i,2}^{\ell},\cdots,\RV{e}_{i,N}^{\ell}]).
\label{nea}
\end{equation}

By combining Node-Node, Edge-Node, and Node-Edge Attention, the proposed block establishes bidirectional information exchange between node and edge representations. This design enables \ac{mdgam} to refine both node states and pairwise relations, including non-Euclidean or asymmetric transition costs, and provides graph representations for the task and communication decoders.

\subsection{Encoder for graph embedding}

As illustrated in Fig.~\ref{fig:main model}, \ac{mdgam} follows an encoder-multi-decoder architecture. The encoder is responsible for transforming the local observation of each agent into graph embeddings that can be used by the subsequent task and communication decoders. For agent $m$, the local observation is organized as a graph containing two types of information: a task-related part, which includes the current agent, its depot, and candidate task nodes, and a communication-related part, which includes neighboring agents reachable through the current communication graph. This construction allows the encoder to jointly represent task-execution context and inter-agent coordination context within a unified graph.

Since different node types have different semantics, their raw features are first projected by type-specific initial embedding layers. We denote the input node features by $(\RV{h}_{\rm own}^{\rm init}, \RV{h}_{\rm dep}^{\rm init}, \RV{h}_{\rm task}^{\rm init}, \RV{h}_{\rm agent}^{\rm init})$, corresponding to the current agent itself, its depot, candidate tasks, and neighboring agents, respectively. After passing through the corresponding embedding layers, the initial node representations are obtained as
\begin{equation}
  \RV{h}^0=[\RV{h}_{\rm own}^0,\RV{h}_{\rm dep}^0,\RV{h}_{\rm task}^0,\RV{h}_{\rm agent}^0] \triangleq [\RV{h}_1^0,\RV{h}_2^0,\ldots,\RV{h}_{N'}^0],
\end{equation}
where $\RV{h}_i^0 \in \mathbb{R}^{d_{\rm m}}$, $i=1,2,\ldots,N'$, and the embedding dimension is set to $d_{\rm m}=128$. 
Meanwhile, edge features $\RV{e}_{ij}^{\rm init}$, $i,j\in\{1,2,\ldots,N'\}$, are introduced to describe the relation between node $i$ and node $j$. In the experiments, they are instantiated as pairwise distances, while the same interface can also accept more general transition-cost inputs. These edge features are projected by a linear layer to obtain the initial edge representations $\RV{e}_{ij}^0$, which participate in the subsequent attention computation.

The initial node and edge representations are then updated by stacking multiple encoder layers. Each encoder layer contains an attention module followed by a feed-forward module \cite{VasShaPar:C17}. In the attention module, Node-Node, Edge-Node, and Node-Edge Attention are sequentially applied to refine node and edge representations. In all \ac{mha} operations, the number of heads is set to 8, and the dimension of each head is $d_{\rm q}=d_{\rm k}=d_{\rm v}=16$. The feed-forward module is applied independently to each node or edge representation and consists of two linear layers with a ReLU activation, with hidden dimension $d_{\rm ff}=256$. Residual connections \cite{HeZhaRen:C16} are applied to both the attention module and the feed-forward module. Moreover, since the size of the local graph changes during sequential decision-making, layer normalization \cite{BaKirHin:A16} is adopted after each encoder layer to stabilize representation learning. We use $\operatorname{Norm}(\cdot)$ to denote layer normalization, and $\overline{\operatorname{FF}}(\cdot)$ to denote the feed-forward module with a residual connection. Accordingly, the representation update in the $\ell$th encoder layer is given by
{
\begin{align}
\hat{\RV{h}}_{i}^{\ell}&=\operatorname{MHA}(\RV{h}_{i}^{\ell-1}, [\RV{h}_1^{\ell-1},\RV{h}_2^{\ell-1},\cdots,\RV{h}_{N'}^{\ell-1}])+\RV{h}_{i}^{\ell-1},\\
\tilde{\RV{e}}_{i,j}^{\ell}&=\operatorname{MHA}(\RV{e}_{ij}^{\ell-1}, \operatorname{PE}(\hat{\RV{h}}_i^{\ell},\hat{\RV{h}}_j^{\ell}))+\RV{e}_{ij}^{\ell-1},\\
\tilde{\RV{h}}_{i}^{\ell}&=\operatorname{MHA}(\hat{\RV{h}}_{i}^{\ell}, [\tilde{\RV{e}}_{i,1}^{\ell},\tilde{\RV{e}}_{i,2}^{\ell},\cdots,\tilde{\RV{e}}_{i,N'}^{\ell}])+\hat{\RV{h}}_{i}^{\ell},\\
\RV{e}_{i,j}^{\ell} &= \operatorname{Norm}(\overline{\operatorname{FF}}(\operatorname{Norm}(\tilde{\RV{e}}_{i,j}^{\ell}))),\\
\RV{h}_{i}^{\ell} &= \operatorname{Norm}(\overline{\operatorname{FF}}(\operatorname{Norm}(\tilde{\RV{h}}_{i}^{\ell}))).
\end{align}}%

After $L$ encoder layers, the resulting $\{\RV{h}_i^L\}$  are used as the final graph node representations. These embeddings summarize the current task-execution context and the available communication context, and are passed to the decoders for task selection and communication-message generation.

\subsection{Decoders for decision and communication}
The encoder provides node and edge representations that summarize the current agent's local observation context. Based on these embeddings, \ac{mdgam} employs two lightweight decoders with different roles. The task decoder produces the probability distribution over the next selectable node, while the communication decoder generates a compact message vector to be shared with connected neighboring agents.

For task selection, the task decoder uses the current decision context as the query and attends to the encoded node representations. The initial decision context $\RV{h}_{\rm c}^0$ is initialized by the encoded representation of the current agent. The task decoder updates this context through $K_1$ decoder layers:
{
\begin{align}
&\hat{\RV{h}}_{\rm c}^k=\operatorname{MHA}({\RV{h}}_{\rm c}^{k-1}, [\RV{h}_1^{L},\RV{h}_2^{L},\cdots,\RV{h}_{N'}^{L}])+\RV{h}_{\rm c}^{k-1},\label{den2n}\\
&{\RV{h}}_{\rm c}^k ={\overline{\operatorname{FF}}}(\hat{\RV{h}}_{\rm c}^k), k=1,2,\ldots,K_1;\\
&P_{\V{\theta}}(a_m^i|\V{o}_m^i)=\operatorname{Output}({\RV{h}}_{\rm c}^{K_1}, [\RV{h}_1^{L},\RV{h}_2^{L},\cdots,\RV{h}_{N'}^{L}], \RV{m}_{\rm c}).
\end{align}}%

The operator $\operatorname{Output}(\cdot)$ is implemented by a single-head attention layer that scores each candidate node. Specifically, the final decision context is projected as the query, and each encoded node representation is projected as a key:
\begin{equation}
{\RV{q}}_{\rm c}=\M{W}^{\rm q}_{\rm out}{\RV{h}}_{\rm c}^{K_1},\quad 
{\RV{k}}_{j}=\M{W}^{\rm k}_{\rm out}\RV{h}_{j}^{L}.
\label{outpre}
\end{equation}
The selection probability of each candidate node is then computed as
{
\begin{align}
&u_j=\begin{cases}
\alpha\cdot\operatorname{tanh}\left(\frac{{\RV{q}}_{\rm c}^{\rm T}{\RV{k}}_{j}}{\sqrt{d_{\rm m}}}\right),&\text{if node $j$ is feasible}\\
-\infty,&\text{otherwise}
\end{cases}\\
&P_{\V{\theta}}(a_m^i=j|\V{o}_m^i)=\operatorname{softmax}_j(u_1,u_2,\ldots,u_{N'}).
\end{align}}%
The mask $\RV{m}_{\rm c}$ is applied to set infeasible nodes to $-\infty$, ensuring that the output distribution assigns probability only to selectable tasks or the depot. During training, the next node is sampled from $P_{\V{\theta}}(a_m^i|\V{o}_m^i)$ to encourage exploration over feasible decisions. During validation and testing, greedy decoding is used, where the feasible node with the highest probability is selected at each decision step. This single-rollout evaluation is consistent with decentralized execution, because repeated trials or post-search refinement are incompatible with practical requirements.

The communication decoder follows a similar attention-based structure, but its output is a continuous message vector rather than a task-selection distribution. Let $\RV{c}^0$ be initialized by the encoded representation of the current agent. For $k=1,2,\ldots,K_2$, the communication context is updated as
{
\begin{align}
&\hat{\RV{c}}^k=\operatorname{MHA}({\RV{c}}^{k-1}, [\RV{h}_1^{L},\RV{h}_2^{L},\cdots,\RV{h}_{N'}^{L}])+\RV{c}^{k-1},\\
&{\RV{c}}^k=\overline{\operatorname{FF}}(\hat{\RV{c}}^k).
\end{align}}%
The final communication message generated by agent $m$ at decision step $i$ is then obtained through a linear projection:
\begin{equation}
\RV{c}_m^i=\M{W}_{\rm com}\RV{c}^{K_2}+\V{b}_{\rm com},
\end{equation}
where $\M{W}_{\rm com}\in\mathbb{R}^{d_{\rm com}\times d_{\rm m}}$, $\V{b}_{\rm com}\in\mathbb{R}^{d_{\rm com}}$, and $d_{\rm com}$ determines the communication-message dimension. The generated message $\RV{c}_m^i$ is used in subsequent local observations of communication-reachable agents.

\section{Training Algorithms}

This section presents the training algorithms for optimizing the proposed \ac{mdgam}. First, we describe a centralized critic based on the extended graph attention mechanism for actor-critic training baselines. We then introduce the proposed \ac{grmapg}, which directly optimizes the terminal task-completion objective through group-relative evaluation without requiring an additional critic network.

\subsection{Actor-Critic Training with an Attention-Based Critic}
\label{subsec:actor_critic}
Actor-critic algorithms, such as \ac{maa2c} \cite{LowWuTam:C17} and \ac{mappo} \cite{YuVelVin:C22}, introduce a trainable critic to estimate state values and provide advantages for actor optimization. Under the \ac{ctde} paradigm, distributed actors make decisions from local observations, while a centralized critic can access global state information during training. In this subsection, the attention-based critic specifically refers to a centralized critic built with the extended graph attention mechanism.

As the basis of policy-gradient algorithms, REINFORCE \cite{Wil:J92} provides a gradient estimator for optimizing a parameterized policy. For a problem instance $\V{g}$, the terminal-reward objective in \eqref{policy_objective} can be written as
\begin{equation}
\mathcal{L}(\V{\theta}|\V{g})
=
\mathbb{E}_{P_{\V{\theta}}(\V{\sigma}|\V{g})}
\left[R(\V{\sigma},\V{g})\right].
\end{equation}
Its gradient is estimated from sampled trajectories as
\begin{equation}
\nabla_{\V{\theta}}\mathcal{L}(\V{\theta}|\V{g})=\mathbb{E}_{P_{\V{\theta}}(\V{\sigma}|\V{g})}\left[\left(R(\V{\sigma},\V{g})-b(\V{g})\right)\nabla_{\V{\theta}}\log  P_{\V{\theta}}(\V{\sigma}|\V{g})\right].
\label{pgloss}
\end{equation}
where the baseline function $b(\V{g})$ reduces estimation variance without changing the expected gradient. Actor-critic methods replace this trajectory-level baseline with a learned value function that estimates the expected return at each decision step.

To support actor-critic training for the graph-structured task-planning problem, we design a centralized critic based on the extended graph attention mechanism. This critic can be integrated with actor-critic algorithms based on state-value estimation, such as \ac{maa2c} and \ac{mappo}. It consists of an extended graph attention encoder and a value head. Similar to the encoder of \ac{mdgam}, the critic encoder processes the global task graph and outputs context, task-node, and agent-node embeddings. At decision step $i$, these encoder outputs are denoted by $\RV{z}_{\rm ctx}^{i}$, $\{\RV{z}_{n}^{i}\}_{n\in{\Set T}}$, and $\{\RV{z}_{m}^{i}\}_{m\in{\Set R}}$, respectively, and are aggregated into the graph-level representation
\begin{equation}
\begin{aligned}
\RV{z}_{\rm g}^{i}=\big[
&\RV{z}_{\rm ctx}^{i};
\operatorname{mean}_{n\in{\Set T}}\RV{z}_{n}^{i};
\operatorname{max}_{n\in{\Set T}}\RV{z}_{n}^{i};\\
&\operatorname{mean}_{m\in{\Set R}}\RV{z}_{m}^{i};
\operatorname{max}_{m\in{\Set R}}\RV{z}_{m}^{i}
\big].
\end{aligned}
\label{critic_graph_embedding}
\end{equation}
The semicolons in \eqref{critic_graph_embedding} denote concatenation. A multilayer-perceptron value head then maps the graph-level embedding to the estimated global state value:
\begin{equation}
V_{\V{\phi}}(\V{s}^{i})=
f_{\V{\phi}}^{\rm V}(\RV{z}_{\rm g}^{i}),
\label{attention_critic_value}
\end{equation}
where $\V{\phi}$ denotes all parameters of the centralized critic, including those of the encoder and value head. Taking \ac{mappo} as a representative example, \ac{mdgam} parameterizes the distributed actor policy, while the proposed centralized critic evaluates the global task-planning state during training. Using an auxiliary stepwise reward $R'(\V{s}^{i},\V{a}^{i})$, the temporal-difference residual and generalized advantage estimate are computed as
\begin{align}
\delta^{i}
&=R'(\V{s}^{i},\V{a}^{i})
+\gamma V_{\V{\phi}}(\V{s}^{i+1})
-V_{\V{\phi}}(\V{s}^{i}),\\
\hat{A}^{i}
&=\sum_{j=0}(\gamma\lambda)^{j}\delta^{i+j},
\label{mappo_advantage}
\end{align}
where the summation terminates at the end of the episode. The critic parameters are optimized using the empirical return $\hat{G}^{i}$:
\begin{equation}
\mathcal{L}_{\rm V}(\V{\phi})
=
\mathbb{E}\left[
\left(V_{\V{\phi}}(\V{s}^{i})-\hat{G}^{i}\right)^2
\right].
\label{critic_loss}
\end{equation}

Let $\V{\theta}_{\rm old}$ denote the behavior-policy parameters used for trajectory sampling. For action $a_m^i$, the probability ratio between the updated and behavior policies is
\begin{equation}
\rho_m^i(\V{\theta})
=
\frac{P_{\V{\theta}}(a_m^i|\V{o}_m^i)}
{P_{\V{\theta}_{\rm old}}(a_m^i|\V{o}_m^i)}.
\label{mappo_ratio}
\end{equation}
\Ac{mappo} updates the shared actor by maximizing the clipped objective. The clipped ratio and actor objective are given by
\begin{align}
\bar{\rho}_m^i(\V{\theta})
&=
\operatorname{clip}\left(\rho_m^i(\V{\theta}),1-\epsilon,1+\epsilon\right),
\label{mappo_clipped_ratio}\\
\mathcal{L}_{\rm A}(\V{\theta})
&=
\mathbb{E}\left[
\sum_{m=1}^{M}\sum_i
\min\left(
\rho_m^i(\V{\theta})\hat{A}^{i},
\bar{\rho}_m^i(\V{\theta})\hat{A}^{i}
\right)
\right].
\label{mappo_obj}
\end{align}

Despite the graph-aware global value estimates provided by the attention-based critic, actor-critic training has two limitations in the task-planning problem considered in this paper. First, the final reward is available only after all agents have completed their task-planning trajectories, making reliable stepwise reward assignment difficult. Second, the additional critic increases the number of trainable parameters and introduces value-estimation errors into policy optimization. To address these limitations, the next subsection presents the critic-free \ac{grmapg} algorithm.

\subsection{Group relative multi-agent policy gradient}
\label{subsec:grmapg}

The proposed \ac{grmapg} algorithm is designed to optimize the  policy objective in \eqref{policy_objective} without introducing an additional critic network.
Its key idea is to evaluate each sampled multi-agent trajectory relative to a group of peer trajectories generated from equivalent task-planning instances \cite{KimParPar:C22,ShaWanZhu:A24}.
For a problem instance $\V{g}$, we construct a group
\begin{equation}
{\Set G}(\V{g})=\{\V{g}^{1},\V{g}^{2},\ldots,\V{g}^{K}\},
\end{equation}
where the group members are obtained from the same original instance through equivalent transformations.
These transformations change certain attributes of the original instance while preserving the theoretical optimal objective.
Examples include rotating the coordinates of all agents and tasks, exchanging task types together with the corresponding agent capabilities, and adaptively scaling temporal parameters and agent speeds so that travel times, service times, and time windows remain consistent under the same time scale.
Such operations are regarded as equivalent transformations induced by the symmetry of the original problem.
In addition, during training, a sampling strategy is used to decode node selections, which further introduces solution-level symmetry.
The combination of problem symmetry and solution symmetry enables the policy network to generate different solutions for a group of equivalent instances, thereby providing mutual supervision among group members.
For each group member $\V{g}^{k}$, the shared policy generates a joint task sequence $\V{\sigma}^{k}\sim P_{\V{\theta}}(\cdot|\V{g}^{k})$, and its terminal reward is denoted by
\begin{equation}
R_k=R(\V{\sigma}^{k},\V{g}^{k}),\quad k=1,2,\ldots,K .
\end{equation}
Instead of learning a value function, \ac{grmapg} uses the rewards of the other group members to construct a leave-one-out baseline:
\begin{equation}
b_k=\frac{1}{K-1}\sum_{\substack{j=1\\j\neq k}}^{K}R_j .
\label{grmapg_baseline}
\end{equation}
The corresponding group-relative advantage is then defined as
\begin{equation}
\hat{A}_k=R_k-b_k .
\label{grmapg_advantage}
\end{equation}
Because the leave-one-out average excludes $R_k$, the \ac{grmapg} baseline $b_k$ is determined only by the rewards of the other group members. Here, the trajectories of different group members are sampled independently conditioned on the transformed instances, the return signals used in $b_k$ are treated with stop-gradient, and the random transformations used to construct the group are independent of the actions generated for $\V{g}^{k}$. Let $\V{\sigma}^{-k}$ denote the set of trajectories of all group members except $\V{\sigma}^{k}$. Under these conditions, $b_k$ is conditionally independent of the current trajectory $\V{\sigma}^{k}$ and satisfies the following property:
\begin{equation}
\begin{aligned}
&\mathbb{E}_{\V{\sigma}^{1:K}}\Bigl[
b_k\nabla_{\V{\theta}}\log P_{\V{\theta}}(\V{\sigma}^{k}|\V{g}^{k})
\Bigr] \\
&\quad =
\mathbb{E}_{\V{\sigma}^{-k}}\Bigl[
b_k\,\mathbb{E}_{\V{\sigma}^{k}}
\bigl[\nabla_{\V{\theta}}\log P_{\V{\theta}}(\V{\sigma}^{k}|\V{g}^{k})\bigr]
\Bigr]
=0 .
\end{aligned}
\label{grmapg_baseline_unbiased}
\end{equation}
Thus, introducing the \ac{grmapg} baseline preserves the unbiasedness of the gradient estimate.

Finally, the gradient estimator of \ac{grmapg} is given by
\begin{equation}
\nabla_{\V{\theta}}\mathcal{L}(\V{\theta}|\V{g})
=
\frac{1}{K}\sum_{k=1}^{K}
\hat{A}_k
\nabla_{\V{\theta}}\log P_{\V{\theta}}(\V{\sigma}^{k}|\V{g}^{k}).
\label{grmapg_gradient}
\end{equation}
According to the trajectory factorization in Section~\ref{subsec:decpomdp}, $\log P_{\V{\theta}}(\V{\sigma}^{k}|\V{g}^{k})$ is obtained by summing the log-probabilities of all decentralized task-selection decisions made by the agents.
In each training iteration, a batch of problem instances is sampled, and a group ${\Set G}(\V{g})$ is constructed for each instance.
\Ac{mdgam} is then invoked to solve each group member through distributed sequential decision-making, thereby generating the corresponding multi-agent trajectories.
After the terminal rewards are evaluated, the \ac{grmapg} baselines and group-relative advantages are computed, and the shared policy parameters are updated using \eqref{grmapg_gradient}.
This critic-free training scheme directly optimizes the terminal task-completion objective, avoids handcrafted stepwise reward decomposition, and introduces no additional trainable parameters beyond the policy network.
The overall training procedure is summarized in Algorithm~\ref{alg:grmapg}.

\begin{algorithm}[t]
\SetAlgoLined
\KwIn{Training distribution $\mathcal{D}$, policy $P_{\V{\theta}}$, group size $K$, batch size $B$, learning rate $\eta$}
\KwOut{Trained policy parameters $\V{\theta}$}
Initialize policy parameters $\V{\theta}$\;
\While{not converged}{
    Sample a batch of instances $\{\V{g}_i\}_{i=1}^{B}\sim\mathcal{D}$\;
    Initialize the gradient accumulator $\V{d}\leftarrow \V{0}$\;
    \For{$i=1$ \KwTo $B$}{
        Construct ${\Set G}(\V{g}_i)=\{\V{g}_i^{1},\ldots,\V{g}_i^{K}\}$ by equivalent transformations\;
        \For{$k=1$ \KwTo $K$}{
            Solve $\V{g}_i^{k}$ with current $P_{\V{\theta}}$\;
            Record  $\V{\sigma}^{k}$ and $L_i^{k}=\log P_{\V{\theta}}(\V{\sigma}^{k}|\V{g}_i^{k})$\;
            Evaluate the reward $R_i^{k}=R(\V{\sigma}^{k},\V{g}_i^{k})$\;
        }
        \For{$k=1$ \KwTo $K$}{
            Compute the baseline $b_{i,k}$ using \eqref{grmapg_baseline}\;
            Compute the advantage $\hat{A}_{i,k}$ using \eqref{grmapg_advantage}\;
            $\V{d}\leftarrow \V{d}+\hat{A}_{i,k}\nabla_{\V{\theta}}L_i^{k}$\;
        }
    }
    Update $\V{\theta}$ using $\V{d}/(BK)$ with learning rate $\eta$\;
}
\caption{Group Relative Multi-Agent Policy Gradient}
\label{alg:grmapg}
\end{algorithm}

The ablation experiments in Section~\ref{subsec:ablation} compare \ac{grmapg}, \ac{mappo}, and \ac{maa2c} using the same \ac{mdgam} policy model. The results show that \ac{grmapg} achieves better training efficiency and convergence performance than these two conventional \ac{marl} algorithms.

\section{Experiments}
We evaluate the proposed method from four perspectives: task-completion performance, ablation analysis, complexity, and generalization capability. The experiments are conducted under different problem scales and communication ranges, with the number of completed tasks used as the primary performance metric. A normalized instance generation scheme is adopted. For each instance, $N$ tasks and $M$ agents are randomly initialized in a square region, and the task time windows, service durations, and agent velocities are generated within prescribed ranges. We set the number of task types to $C=2$ in the experiments. Each task is assigned one of the two types, and the proportion of either type is randomly set between $0.4$ and $0.6$. The agent capabilities are also randomly assigned to reflect heterogeneous task-execution abilities. Specifically, agents that can execute only the first task type, agents that can execute only the second task type, and agents with both capabilities each account for approximately one third of the agent population.

\textbf{Model and Training Settings:} In the experiments, we use an encoder with $L=4$ layers and set the numbers of layers in the task decoder and the communication decoder to $K_1=K_2=1$, respectively. The embedding dimension is set to $d_{\rm m}=128$, and all \ac{mha} modules use $8$ heads with $d_{\rm q}=d_{\rm k}=d_{\rm v}=16$ for each head. The hidden dimension of the feed-forward module is set to $d_{\rm ff}=256$, and the communication-message dimension is set to $d_{\rm com}=128$. The model is trained using \ac{grmapg} with the group size set to $K=4$. We use the Adam optimizer \cite{KinBa:C15} with a batch size of $64$, and each epoch contains $200$ randomly generated batches. For each fixed instance configuration, the model is trained for $20$ epochs, with the learning rate exponentially decayed from $\eta=10^{-4}$ to $\eta=10^{-5}$. The training process is conducted on four NVIDIA RTX 3090 GPUs. As a reference for training time, the complete training process takes approximately $27$ hours under the setting $N=100$ and $M=7$. When the problem scale changes, loading a pretrained model from another scale can substantially reduce the training time compared with training from scratch, and convergence is typically achieved within about $8$ epochs. During training, task selections are sampled from the policy distribution for exploration. In validation and testing, greedy decoding is used to evaluate the capability of the policy model to make one-shot decisions on problem instances. Validation is performed on a single GPU with the batch size set to $1$.

The detailed experimental results are presented in the following four subsections. To facilitate reproducibility, the source code of the experiments is publicly available at \url{https://anonymous.4open.science/r/mdgam_code_anonymous-F175}. The released repository includes the implementations of the evaluated models, as well as the training and testing scripts used in this paper.

\subsection{Main results}

We evaluate the main task-completion performance under three problem scales, namely $(N=50, M=4)$, $(N=100, M=7)$, and $(N=150, M=10)$, corresponding to small-, medium-, and large-scale instances, respectively. For each scale, three communication ranges, i.e., $r=0.2$, $0.4$, and $0.6$, are considered to represent strict, moderate, and relaxed communication constraints. The proposed method is compared with representative heuristic and \ac{rl}-based baseline algorithms. The baseline algorithms are summarized as follows:
  \begin{itemize}
      \item \textbf{Auction-based heuristics:} We consider \ac{cbba} \cite{BruChoHow:C08}, PI-maxAss \cite{TurMenSch:J17}, and EEPI \cite{WanLiuQiu:J23}. To accommodate the dynamic communication topology considered in this paper, each method performs distributed planning among the agents within the current communication-connected component. Replanning is triggered when the decision-making agent observes a newly connected agent joining its component.
      \item \textbf{\Ac{rl}-based methods:} We compare with DL-DRL \cite{MaoWuFan:J24} and CAM \cite{PauCho:J24}. Since DL-DRL follows a centralized task-assignment and sequence-planning structure, it is trained with global information and evaluated by centralized decision-making within each communication-connected component. CAM is trained under the same distributed setting as the proposed method, where unencoded agent attributes are transmitted during the communication phase. 
For both learning-based baselines, the model sizes and key hyperparameters are kept as consistent as possible with those of the proposed method.
  \end{itemize}

\begin{table*}[ht]
\caption{Number of completed tasks achieved by the proposed method and baseline algorithms under different problem scales and communication ranges. The best-performing method under each setting is highlighted in bold.}
\label{table:main results}
\begin{center}
\begin{small}
\begin{tabular}{lc|ccc|ccc|ccc}
\toprule
\multirow{2}{*}{Method} & \multirow{2}{*}{Type} & \multicolumn{3}{c|}{{50 tasks, 4 agents}} & \multicolumn{3}{c|}{{100 tasks, 7 agents}} & \multicolumn{3}{c}{{150 tasks, 10 agents}}  \\
 & & $r=0.2$ & $r=0.4$ & $r=0.6$ & $r=0.2$ & $r=0.4$ & $r=0.6$ & $r=0.2$ & $r=0.4$ & $r=0.6$\\
\midrule
CBBA & Heuristic &32.11&33.70&33.93&66.85&69.87&70.66&105.68& 107.79&109.91\\
PI-maxAss & Heuristic &36.41&37.98&38.49&75.41&79.61&80.19&117.96&124.34&124.70 \\
EEPI & Heuristic &35.35&36.81&37.27&74.18&77.85&78.34&116.13&121.76&122.15 \\
DL-DRL & RL &30.04&31.92&33.20&61.88&67.05&68.90&91.70&101.57&106.75 \\
CAM & RL &36.27&37.75&38.33&74.54&79.91&81.10&117.65&125.37& 126.48\\
Ours & RL &\textbf{36.64}&\textbf{38.83}&\textbf{39.58}&\textbf{77.42}&\textbf{82.90}&\textbf{84.15}&\textbf{123.60}&\textbf{131.72}&\textbf{133.11}\\
\bottomrule
\end{tabular}
\end{small}
\end{center}
\end{table*}

\begin{figure*}[ht]
\begin{minipage}[c]{1.0\linewidth}
  \centering
  \centerline{\includegraphics[width=0.9\textwidth]{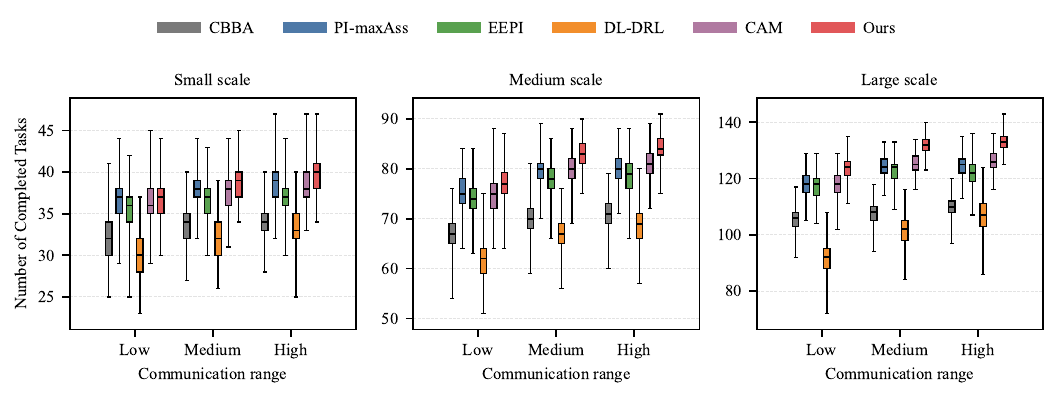}}
\end{minipage}
\caption{Box plots of task-completion results under different problem scales and communication ranges. For each method, the whiskers indicate the minimum and maximum numbers of completed tasks on the test dataset, while the lower edge, middle line, and upper edge of the box represent the first quartile, median, and third quartile, respectively.}
\label{fig:main results}
\end{figure*}

For each scale, we construct a dataset containing 1000 instances and evaluate the task-completion performance of the proposed method and all baseline algorithms. The average number of completed tasks achieved by each method is reported in Table~\ref{table:main results}, and the distribution of the task-completion results is illustrated using box plots in Fig.~\ref{fig:main results}.



As shown in Table~\ref{table:main results}, the proposed method achieves the highest average number of completed tasks in all nine combinations of problem scale and communication range. Compared with PI-maxAss, which is the strongest heuristic baseline in most settings, the proposed method obtains increasingly larger gains as the problem scale grows. Specifically, when $r=0.4$, the improvements over PI-maxAss are $2.24\%$, $4.13\%$, and $5.94\%$ on the small-, medium-, and large-scale datasets, respectively. This scaling trend is also illustrated by the box plots in Fig.~\ref{fig:main results}, where the distribution of completed-task numbers achieved by the proposed method is consistently shifted upward and the separation from the baseline algorithms becomes more evident on larger-scale instances.

The proposed method also outperforms the learning-based baselines DL-DRL and CAM under all evaluated settings. When $r=0.4$, the proposed method improves the number of completed tasks over CAM by $2.86\%$, $3.74\%$, and $5.06\%$ on the small-, medium-, and large-scale datasets, respectively. These results show that, under the same distributed decision-making framework, the proposed policy architecture provides stronger cooperative task-scheduling performance than the existing learning-based policies. In addition, increasing the communication range generally improves the performance of all methods, since a larger communication range provides each agent with more neighboring information for coordinated decision-making. Even under the strict communication setting $r=0.2$, the proposed method still achieves the best task-completion performance, demonstrating its effectiveness under limited information exchange.

\subsection{Ablation}
\label{subsec:ablation}

We conduct ablation studies to separately evaluate the contributions of the proposed \ac{mdgam} architecture and the \ac{grmapg} training algorithm.

\textbf{Model Ablation:} We evaluate the architectural contribution of \ac{mdgam} from two aspects. First, to examine the effect of the communication decoder, we construct a single-decoder variant by removing the communication decoder from \ac{mdgam}; in this variant, agents make task-selection decisions without generating learned communication messages. Second, to evaluate the graph representation module, we compare \ac{mdgam} with two representative graph neural network architectures, namely \ac{am} \cite{KoovanWel:C19} and \ac{gcn} \cite{JosCapRouLau:J22}. The representation dimensions of \ac{am} and \ac{gcn} are set to be the same as those of \ac{mdgam}, and their total numbers of trainable parameters are kept at a similar scale. All model variants are trained using the same \ac{grmapg} algorithm and identical training settings, and are evaluated under the medium-scale setting with $N=100$ and $M=7$.

\textbf{Training Algorithm Ablation:} We compare the proposed \ac{grmapg} with two conventional \ac{marl} algorithms, \ac{mappo} and \ac{maa2c}, under the same distributed multi-agent task-planning framework. All three training algorithms use the same data settings, including the same instance distribution, batch size, number of epochs, and learning rate. \Ac{mappo} and \ac{maa2c} also use the grouped input construction with $K=4$ as a data-augmentation strategy. Therefore, the three algorithms use exactly the same number of instances and policy-inference calls, which ensures a fair comparison. For \ac{mappo} and \ac{maa2c}, the centralized critic described in Section~\ref{subsec:actor_critic} is used for state-value estimation and actor optimization.

\begin{table}[!t]
\caption{Ablation results under the medium-scale setting. The variant ``w/o Comm. Decoder'' denotes \ac{mdgam} with the communication decoder removed.}
\label{table:ablation}
\begin{center}
\begin{small}
\begin{tabular}{l|ccc}
\toprule
\multirow{2}{*}{Method} &  \multicolumn{3}{c}{{100 tasks, 7 agents}} \\
& $r=0.6$ & $r=0.4$ & $r=0.2$\\
\midrule
{Original}&\textbf{84.15} &\textbf{82.90}& \textbf{77.42} \\
\midrule
\multicolumn{4}{l}{\textit{Model Ablation}}\\
\midrule
{w/o Comm. Decoder} & 82.80 & 81.63 & 76.65\\
{AM} & 81.52&80.45&75.53\\
{GCN} &81.75 &81.08 &76.11 \\
\midrule
\multicolumn{4}{l}{\textit{Training Algorithm Ablation}}\\
\midrule
{MAPPO} &83.17&81.99&76.76\\
{MAA2C} &82.52&81.59&76.34 \\
\bottomrule
\end{tabular}
\end{small}
\end{center}
\end{table}
\begin{figure}[!t]
\begin{minipage}[c]{1.0\linewidth}
  \centering
  \centerline{\includegraphics[width=0.95\textwidth]{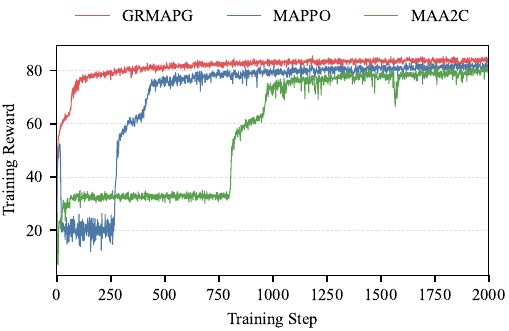}}
\end{minipage}
\caption{Training reward curves of different policy-gradient algorithms under the medium-scale setting with $N=100$, $M=7$, and $r=0.6$.}
\label{fig:training curves}
\end{figure}

Table~\ref{table:ablation} reports the ablation results under the medium-scale setting with $N=100$ and $M=7$, covering three communication ranges. The original method, which combines \ac{mdgam} and \ac{grmapg}, achieves the highest average number of completed tasks in all three settings. In the model ablation, \ac{mdgam} consistently outperforms the single-decoder variant, \ac{am}, and \ac{gcn}. Compared with the variant without the communication decoder, \ac{mdgam} increases the average number of completed tasks by $1.35$, $1.27$, and $0.77$ when $r=0.6$, $0.4$, and $0.2$, respectively. Compared with \ac{am}, the corresponding improvements are $2.63$, $2.45$, and $1.89$, while compared with \ac{gcn}, the improvements are $2.40$, $1.82$, and $1.31$. These results verify that both the extended graph attention mechanism and the multi-decoder design improve the capability of the policy model for decentralized task scheduling.

The training-algorithm ablation further verifies the effectiveness of \ac{grmapg}. With the same \ac{mdgam} actor, \ac{grmapg} achieves higher task-completion performance than \ac{mappo} and \ac{maa2c} across all communication ranges in Table~\ref{table:ablation}. These results show that the proposed critic-free policy-gradient estimator provides an additional gain beyond the network architecture. Fig.~\ref{fig:training curves} further compares the training reward curves under the setting $N=100$, $M=7$, and $r=0.6$. \Ac{grmapg} reaches higher rewards more rapidly during training, whereas \ac{mappo} and \ac{maa2c} exhibit slower improvement in the early stage. This result is consistent with the discussion in Section~\ref{subsec:grmapg}: directly optimizing the terminal task-completion objective with group-relative advantages avoids the need to train an additional value function and improves the convergence behavior of policy optimization.

\subsection{Complexity}
We evaluate the computational and communication costs of the proposed distributed decision-making framework and conventional PI-based algorithms. CAM and DL-DRL are not included because, under the distributed setting considered in this paper, they adopt the same sequential decision-making procedure and therefore have similar complexity profiles. The computational cost is mainly evaluated by the total runtime required to solve test instances. PI-maxAss and EEPI are run on one CPU core, while the proposed neural policy is evaluated on a single GPU with the batch size set to $1$, since GPU acceleration is a practical choice for neural-network inference on modern edge-computing platforms. We record the total runtime required by each method to solve 1000 instances under the communication range $r=0.4$, as reported in Table~\ref{Time cost}. Under this tested implementation and hardware setting, the proposed method requires more time than EEPI on the small-scale instances, but its runtime grows much more slowly as the problem scale increases. On the large-scale setting with $150$ tasks and $10$ agents, the proposed method solves 1000 instances in $31$ min $7$ s, corresponding to approximately $1.9$ s per instance. This is about $95$ times faster than PI-maxAss and $5.7$ times faster than EEPI. These results show that the learned policy provides clear inference-time advantages on medium- and large-scale instances, while also achieving higher task-completion performance as shown in Table~\ref{table:main results}.

\begin{table}[h]
\caption{Total runtime for solving 1000 instances with $r=0.4$.}
\label{Time cost}
\begin{center}
\begin{small}
\begin{tabular}{l|ccc}
\toprule
{Method} &  (50,4) & (100,7) & (150,10)\\
\midrule
{PI-maxAss} &1h12m &14h31m  &49h2m \\
{EEPI} &6m37s  &1h13m &2h59m \\
{Ours} &10m38s &21m25s & 31m7s\\
\bottomrule
\end{tabular}
\end{small}
\end{center}
\end{table}

\begin{table}[h]
\caption{Average communication cost per instance with $r=0.4$.}
\label{Communication cost}
\begin{center}
\begin{small}
\begin{tabular}{l|ccc}
\toprule
{Method} &  (50,4) & (100,7) & (150,10)\\
\midrule
{PI-maxAss} &209 &1848 &5272 \\
{EEPI} &243 &1654 & 3905\\
{Ours} &83 &610 & 2149\\
\bottomrule
\end{tabular}
\end{small}
\end{center}
\end{table}

The communication cost is defined as the number of message transmissions among connected agents during the decision-making process. For information exchange between two agents that requires multi-hop propagation, each hop is counted as one message transmission. As shown in Table~\ref{Communication cost}, the proposed method requires fewer message transmissions than both PI-maxAss and EEPI across all evaluated scales. In the large-scale setting, the communication count is reduced from $5272$ to $2149$ compared with PI-maxAss, and from $3905$ to $2149$ compared with EEPI. The communication count of the proposed method still increases with the number of agents, because the current design allows the decision-making agent to receive messages from all agents within its communication-connected component. In practical deployments with stricter communication budgets, this overhead can be further controlled by reducing the message dimension or limiting the number of neighboring agents considered at each decision step.

\subsection{Generalization tests}
This subsection evaluates the generalization capability of the proposed model across different problem settings. Owing to its graph-based architecture, \ac{mdgam} can process different combinations of the number of tasks, the number of agents, and the communication range, denoted by $(N,M,r)$, without modifying the input format or network structure. In each generalization experiment, the model is trained under one setting of $(N,M,r)$ and then evaluated on unseen test instances generated under different settings. Specifically, we vary one of $N$, $M$, and $r$ while keeping the other two factors fixed.

\begin{figure}[t!]
\begin{minipage}[c]{1.0\linewidth}
  \centering
  \centerline{\includegraphics[width=0.9\textwidth]{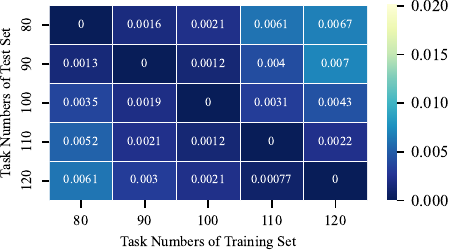}}
\end{minipage}
\caption{Generalization results under different numbers of tasks with $7$ agents and $r=0.4$.}
\label{generalization_heatmap_M}
\end{figure}

\begin{figure}[t!]
\begin{minipage}[c]{1.0\linewidth}
  \centering
  \centerline{\includegraphics[width=0.9\textwidth]{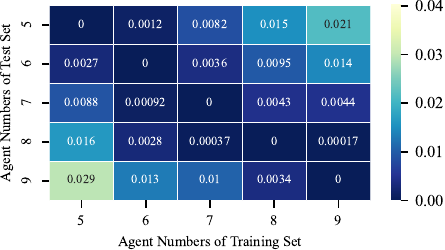}}
\end{minipage}
\caption{Generalization results under different numbers of agents with $100$ tasks and $r=0.4$.}
\label{generalization_heatmap_N}
\end{figure}

\begin{figure}[t!]
\begin{minipage}[c]{1.0\linewidth}
  \centering
  \centerline{\includegraphics[width=0.9\textwidth]{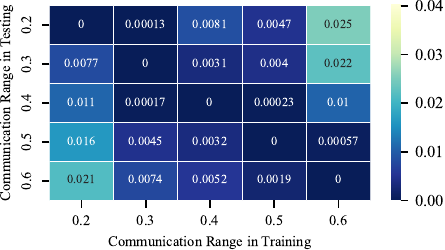}}
\end{minipage}
\caption{Generalization results under different communication ranges with $100$ tasks and $7$ agents.}
\label{generalization_heatmap_C}
\end{figure}

Figs.~\ref{generalization_heatmap_M}, \ref{generalization_heatmap_N}, and \ref{generalization_heatmap_C} present the generalization results with respect to the number of tasks, the number of agents, and the communication range, respectively. In each heatmap, a cell denotes the performance gap between the model trained under the corresponding column setting and the best-performing model on the corresponding row test set. The diagonal entries indicate identical training and test settings, and therefore have a gap of $0$. Overall, the three heatmaps show that the performance gap generally increases as the training and test settings become more different, indicating the effect of distribution mismatch on policy generalization.

Specifically, Fig.~\ref{generalization_heatmap_M} shows that the maximum performance gap among models trained with different numbers of tasks is only $0.7\%$, indicating strong generalization capability with respect to task-number variations. This result suggests that differences in the number of tasks between training and testing have only a limited effect on performance. In Fig.~\ref{generalization_heatmap_N}, the performance degradation is less than $1\%$ when the difference between the training and test numbers of agents is no larger than $2$. When this difference increases to $4$, the degradation becomes more evident, reaching approximately $3\%$. A possible reason is that changing the number of agents may alter the agent density and the level of inter-agent coordination required by the policy.

Fig.~\ref{generalization_heatmap_C} evaluates generalization across communication ranges from the communication-constrained case $r=0.2$ to the communication-relaxed case $r=0.6$. The results show different generalization behaviors under different communication conditions. Models trained under extreme communication settings, i.e., $r=0.2$ and $r=0.6$, show relatively larger performance gaps when transferred to other values of $r$. In contrast, models trained under intermediate communication ranges, i.e., $r=0.3$--$0.5$, maintain favorable performance across a wider range of test settings. Overall, the proposed model demonstrates good generalization capability with respect to variations in task number, agent number, and communication range, with only limited performance degradation under moderate distribution shifts.

\section{Conclusion}
This paper investigated cooperative task scheduling for communication-constrained distributed multi-agent systems, where agents make decisions from local observations while satisfying heterogeneous agent attributes, various task constraints, and limited communication ranges. To address this problem, we formulated the distributed task planning process as a \ac{decpomdp} and proposed an \ac{mdgam}-based neural scheduling framework that jointly generates task-selection decisions and communication messages. The proposed \ac{mdgam} uses an extended graph attention encoder to represent graph-structured local observations with both node and edge features, and employs task and communication decoders to generate decentralized decisions and inter-agent messages. For policy optimization, we developed \ac{grmapg}, a critic-free training algorithm that constructs group-relative advantages without ground truth, reduces trainable parameters, and improves convergence.

Extensive experiments were conducted under different problem scales and communication ranges. The results show that the proposed method achieves higher task-completion performance than representative heuristic methods, including \ac{cbba}, PI-maxAss, and EEPI, as well as learning-based methods such as DL-DRL and CAM. The complexity analysis further shows that the proposed framework reduces both runtime and communication cost compared with PI-based distributed allocation algorithms, especially on large-scale instances. Ablation studies confirm the performance gains brought by the extended graph attention mechanism, the communication decoder, and the \ac{grmapg} training algorithm. Generalization tests further indicate that the learned policy can adapt to variations in task number, agent number, and communication range with limited performance degradation. Future work will consider more dynamic environments, including online task arrivals, uncertain task execution times, and adaptive communication policies under bandwidth or energy constraints.

\nocite{WanYanHua:A26}
\bibliography{example_paper_IEEE}
\bibliographystyle{IEEEtran}

\vfill

\end{document}

%% file: Wgroup-Teaching-Preamble.tex
\usepackage{bm}
\usepackage{upgreek}

\DeclareMathAlphabet{\mathsfbr}{OT1}{cmss}{m}{n}
\SetMathAlphabet{\mathsfbr}{bold}{OT1}{cmss}{bx}{n}
\DeclareRobustCommand{\msf}[1]{%
  \ifcat\noexpand#1\relax\msfgreek{#1}\else\mathsfbr{#1}\fi
}

\makeatletter
\newcommand{\msfgreek}[1]{\csname s\expandafter\@gobble\string#1\endcsname}
\makeatother

\DeclareFontEncoding{LGR}{}{} 
\DeclareSymbolFont{sfgreek}{LGR}{cmss}{m}{n}
\SetSymbolFont{sfgreek}{bold}{LGR}{cmss}{bx}{n}
\DeclareMathSymbol{\salpha}{\mathord}{sfgreek}{`a}
\DeclareMathSymbol{\sbeta}{\mathord}{sfgreek}{`b}
\DeclareMathSymbol{\sgamma}{\mathord}{sfgreek}{`g}
\DeclareMathSymbol{\sdelta}{\mathord}{sfgreek}{`d}
\DeclareMathSymbol{\sepsilon}{\mathord}{sfgreek}{`e}
\DeclareMathSymbol{\szeta}{\mathord}{sfgreek}{`z}
\DeclareMathSymbol{\seta}{\mathord}{sfgreek}{`h}
\DeclareMathSymbol{\stheta}{\mathord}{sfgreek}{`j}
\DeclareMathSymbol{\siota}{\mathord}{sfgreek}{`i}
\DeclareMathSymbol{\skappa}{\mathord}{sfgreek}{`k}
\DeclareMathSymbol{\slambda}{\mathord}{sfgreek}{`l}
\DeclareMathSymbol{\smu}{\mathord}{sfgreek}{`m}
\DeclareMathSymbol{\snu}{\mathord}{sfgreek}{`n}
\DeclareMathSymbol{\sxi}{\mathord}{sfgreek}{`x}
\DeclareMathSymbol{\somicron}{\mathord}{sfgreek}{`o}
\DeclareMathSymbol{\spi}{\mathord}{sfgreek}{`p}
\DeclareMathSymbol{\srho}{\mathord}{sfgreek}{`r}
\DeclareMathSymbol{\ssigma}{\mathord}{sfgreek}{`s}
\DeclareMathSymbol{\stau}{\mathord}{sfgreek}{`t}
\DeclareMathSymbol{\supsilon}{\mathord}{sfgreek}{`u}
\DeclareMathSymbol{\sphi}{\mathord}{sfgreek}{`f}
\DeclareMathSymbol{\schi}{\mathord}{sfgreek}{`q}
\DeclareMathSymbol{\spsi}{\mathord}{sfgreek}{`y}
\DeclareMathSymbol{\somega}{\mathord}{sfgreek}{`w}

\DeclareMathSymbol{\svarsigma}{\mathord}{sfgreek}{`c}

\DeclareMathSymbol{\sGamma}{\mathalpha}{sfgreek}{`G}
\DeclareMathSymbol{\sDelta}{\mathalpha}{sfgreek}{`D}
\DeclareMathSymbol{\sTheta}{\mathalpha}{sfgreek}{`J}
\DeclareMathSymbol{\sLambda}{\mathalpha}{sfgreek}{`L}
\DeclareMathSymbol{\sXi}{\mathalpha}{sfgreek}{`X}
\DeclareMathSymbol{\sPi}{\mathalpha}{sfgreek}{`P}
\DeclareMathSymbol{\sSigma}{\mathalpha}{sfgreek}{`S}
\DeclareMathSymbol{\sUpsilon}{\mathalpha}{sfgreek}{`U}
\DeclareMathSymbol{\sPhi}{\mathalpha}{sfgreek}{`F}
\DeclareMathSymbol{\sPsi}{\mathalpha}{sfgreek}{`Y}
\DeclareMathSymbol{\sOmega}{\mathalpha}{sfgreek}{`W}

\DeclareRobustCommand{\mcal}[1]{%
  \ifcat\noexpand#1\relax\mathnormal{#1}\else\cal{#1}\fi
}
\DeclareRobustCommand{\BM}[1]{%
  \ifcat\noexpand#1\relax\bm{\boldUppercaseItalicGreek{#1}}\else\bm{#1}\fi
}
\makeatletter
\newcommand{\boldUppercaseItalicGreek}[1]{\csname var\expandafter\@gobble\string#1\endcsname}
\makeatother
  
\newcommand{\RV}[1]{\bm{\MakeLowercase{\msf{#1}}}}
\newcommand{\RM}[1]{\bm{\MakeUppercase{\msf{#1}}}}

\newcommand{\V}[1]{\bm{#1}}
\newcommand{\M}[1]{\BM{#1}}
\newcommand{\Set}[1]{\mcal{#1}}